\documentclass[10pt,twocolumn,letterpaper]{article}

\usepackage{iccv}
\usepackage{times}
\usepackage{epsfig}
\usepackage{graphicx}
\usepackage{amsmath}
\usepackage{amssymb}

\usepackage{makecell}
\usepackage{booktabs}
\usepackage{multirow}
\usepackage{array}
\usepackage{courier}

\usepackage[pagebackref=true,breaklinks=true,letterpaper=true,colorlinks,bookmarks=false]{hyperref}

\iccvfinalcopy 

\def\iccvPaperID{3351} 

\ificcvfinal\fi

\newcommand{\wt}{\textcolor[RGB]{0,0,0}}

\begin{document}

\title{Bi-VLGM : Bi-Level Class-Severity-Aware Vision-Language Graph Matching for Text Guided Medical Image Segmentation}

\author{Wenting Chen$^1$, Jie Liu$^1$ and Yixuan Yuan$^2$\thanks{Corresponding authors: Yixuan Yuan (\href{mailto:yxyuan@ee.cuhk.edu.hk}{yxyuan@ee.cuhk.edu.hk})}\\[2mm]
$^1$City University of Hong Kong~~~~$^2$Chinese University of Hong Kong\\ [0.5mm]
}


\maketitle
\ificcvfinal\thispagestyle{empty}\fi

\begin{abstract}

Medical reports with substantial information can be naturally complementary to medical images for computer vision tasks, and the modality gap between vision and language can be solved by vision-language matching (VLM). However, current vision-language models distort the intra-model relation and \wt{mainly include class information in prompt learning that is insufficient for segmentation task.}
In this paper, we introduce a Bi-level class-severity-aware Vision-Language Graph Matching (Bi-VLGM) for text guided medical image segmentation, composed of a word-level VLGM module and a sentence-level VLGM module, to exploit the class-severity-aware relation among visual-textual features. In word-level VLGM, to mitigate the distorted intra-modal relation during VLM, we reformulate VLM as graph matching problem and introduce a vision-language graph matching (VLGM) to exploit the high-order relation among visual-textual features. Then, we perform VLGM between the local features for each class region and class-aware prompts to bridge their gap. In sentence-level VLGM, to provide disease severity information for segmentation task, we introduce a severity-aware prompting to quantify the severity level of retinal lesion, and perform VLGM between the global features and the severity-aware prompts. By exploiting the relation between the local (global) and class (severity) features, the segmentation model can selectively learn the class-aware and severity-aware information to promote performance.
Extensive experiments prove the effectiveness of our method and its superiority to existing methods. Source code is to be released.
  
\end{abstract}

\section{Introduction}
\label{sec:intro}

Medical image segmentation, classifying the pixels of anatomical or pathological regions from background medical images, is an important tool to deliver critical information about the shapes and volumes of these regions. 
Traditional automatic segmentation methods, with pixel-by-pixel supervision from deep learning, generally ignore the semantic information from medical reports that can provide additional supervision signals for diagnosis \cite{monajatipoor2022berthop}. To leverage the semantic information, several text guided segmentation methods \cite{wen2022fleak,Tomar2022TGANet,Dai2018Clinical,Li2022LViT} propose to directly integrate medical reports to medical images.
However, the heterogeneity gap between text and image \cite{cheng2021bridging} hinders the effectiveness of text information, leading to the imperfection of these methods.

\begin{figure}[t]
  \centering
   \includegraphics[width=\linewidth]{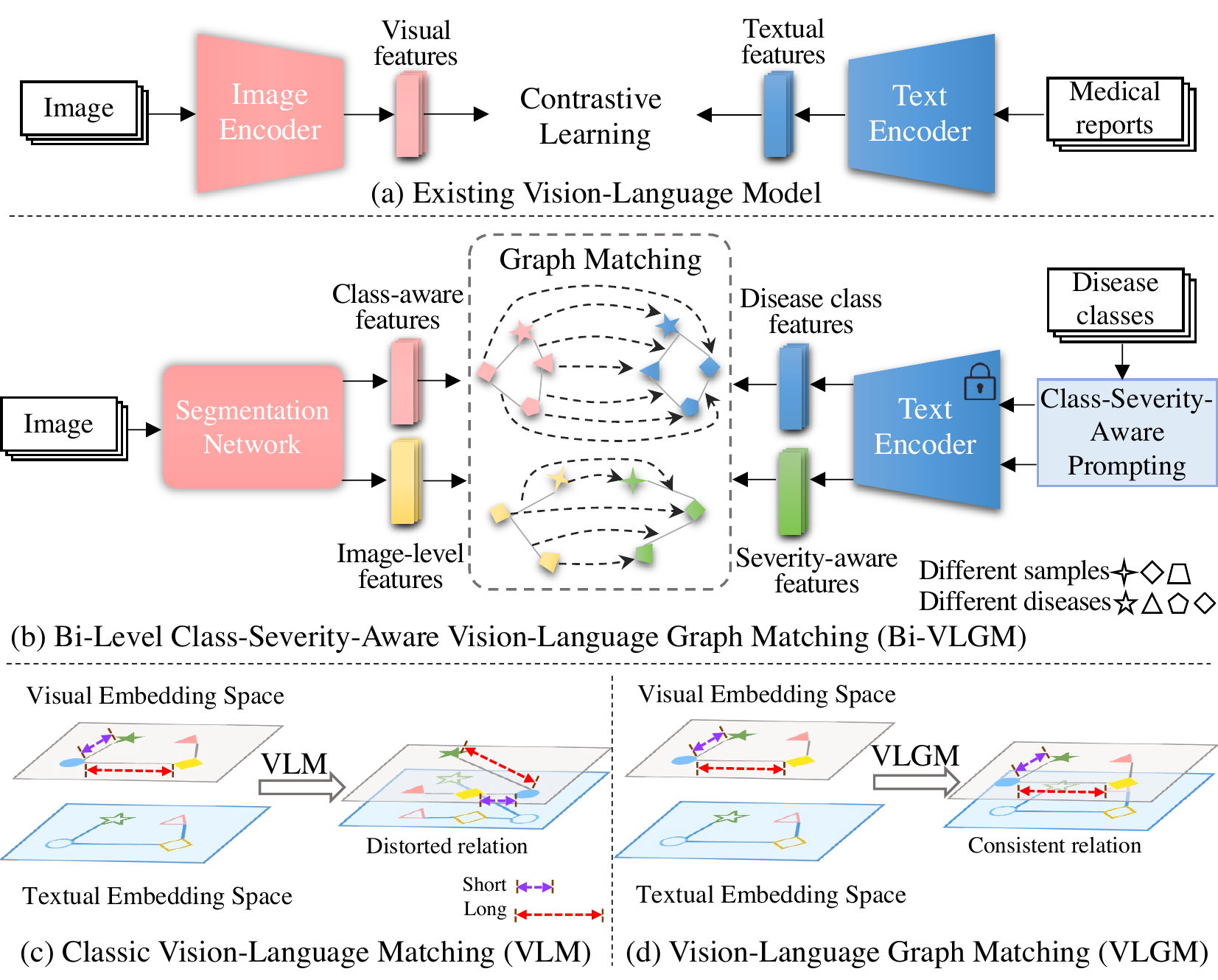}
   \caption{(a) Existing vision-language models apply contrastive learning to perform vision-language matching. (b) The proposed Bi-VLGM aims to match the visual and textual features that include the class-aware and severity-aware information, so as to supervise the learning of the predicted mask. Among the vision-language matching (VLM) methods, we find out that (c) classic VLM distorts the intra-modal relation and (d) the proposed VLGM can remain the relation consistent.}
   \label{fig:overview}
\end{figure}

Recently, the vision-language models are widely investigated and have achieved significant performance in cross-modal tasks. The goal of vision-language models is to jointly learn image and text representations that can be applied to various downstream tasks such as image classification, image caption and visual question answers (VQA) \cite{li2022blip}. As shown in Fig. \ref{fig:overview} (a), the recently proposed vision-language models, like CLIP \cite{radford2021learning} and ALIGN \cite{jia2021scaling}, collect millions of image-text pairs and align visual-textual features by contrastive loss, which bridge the heterogeneity gap between text and image for image classification through vision-language matching (VLM).

Despite the satisfactory performance of vision-language models in classification task, there are two main challenges in the existing vision-language models. Firstly, current vision-language models \cite{shukor2022efficient,huang2022idea,monajatipoor2022berthop} directly align inter-modal features \wt{between vision and language} in point-to-point manner and overlook the relation among the intra-modal features \wt{for vision or language}, which may distort the intra-modal feature space when performing vision-language matching (VLM). As shown in Fig.\ref{fig:overview} (b), after VLM, the distance between the embeddings of star (e.g. zebra) and circle (e.g horse) becomes larger while that for square (e.g. tiger) and circle becomes smaller, making the features of two similar classes further away and those of two dissimilar classes closer.
Such distorted relation would warp the original feature manifold, leading to the misclassification of objects from similar classes.
Instead, encouraging the intra-modal relation consistent contributes to a better feature representation \cite{baek2021exploiting}, as depicted in Fig. \ref{fig:overview} (c). Thus, it is highly recommended to consider the high-order relation among inter-modal features reserving \textbf{the intra-modal relation}.

Another challenge is that existing vision-language models primarily apply the prompts with class information for VLM, ignoring the disease severity. 
These methods \cite{radford2021learning} produce prompts by integrating the class name to the manually designed context that is meaningful to the task, or automatically generate class-specific context combined with the class token \cite{zhou2022learning,zhou2022conditional}. Despite their satisfactory performance on classification task, these methods neglect specific information of diseases (e.g. disease severity) that is useful for segmentation task. When segmenting biomedical objects, there is a significant diversity with their severity level (e.g. size and area) that is useful context information to localize the objects \cite{fan2020pranet,shen2021hrenet}. As a result, the representation learned by these vision-language models may not be optimal for the dense prediction tasks such as segmentation. Thus, it is necessary for prompting engineering to further include the \textbf{disease severity information} for segmentation task.

To overcome the aforementioned challenges, we propose a Bi-level class-severity-aware Vision-Language Graph Matching (Bi-VLGM) for text guided medical image segmentation to exploit the class-severity-aware relation among visual-textual features in word level and sentence level.
Specifically, Bi-VLGM consists of a \textit{word-level VLGM module} and a \textit{sentence-level VLGM module} to model the class and severity relation between visual and textual features. In \textit{word-level VLGM module}, we introduce a class-aware prompting to automatically generate class-aware prompts as class features, and perform alignment of class features and local features for each lesion class. 
To preserve the \textbf{intra-modal relation} during alignment, we innovatively reformulate the VLM to graph matching problem, and introduce a vision-language graph matching (VLGM) to utilize the high-order relation among intra-graph nodes to remain the intra-modal relation consistent. 
In \textit{sentence-level VLGM module}, to provide \textbf{disease severity information}, we propose a severity-aware prompting to quantify lesion severity level with severity-aware prompts as severity features, and perform VLGM between the severity features and image global features integrated with all local features, to explicitly mine their relation. By adopting bi-level VLGM for the ground-truth and predicted masks, respectively, we can bridge the gap between local (global) and class (severity) features, and make the segmentation model to selectively learn the class-aware and severity-aware information, so as to promote the segmentation performance. 
Our contributions can be summarized as follows:

\begin{itemize}
\item We propose a Bi-level class-severity-aware Vision-Language Graph Matching (Bi-VLGM) for text guided medical image segmentation, which performs local-class alignment in word level and global-severity alignment in sentence level to promote segmentation performance.



\item We propose a Vision-Language Graph Matching (VLGM) to reformulate VLM as graph matching problem to remain intra-modal consistent, and introduce a severity-aware prompting to apply the disease severity information for vision-language models. 

\item Extensive experiments on two publicly available medical datasets demonstrate the effectiveness of Bi-VLGM and its superior performance.
\end{itemize}

\begin{figure*}[t]
  \centering
   \includegraphics[width=\linewidth]{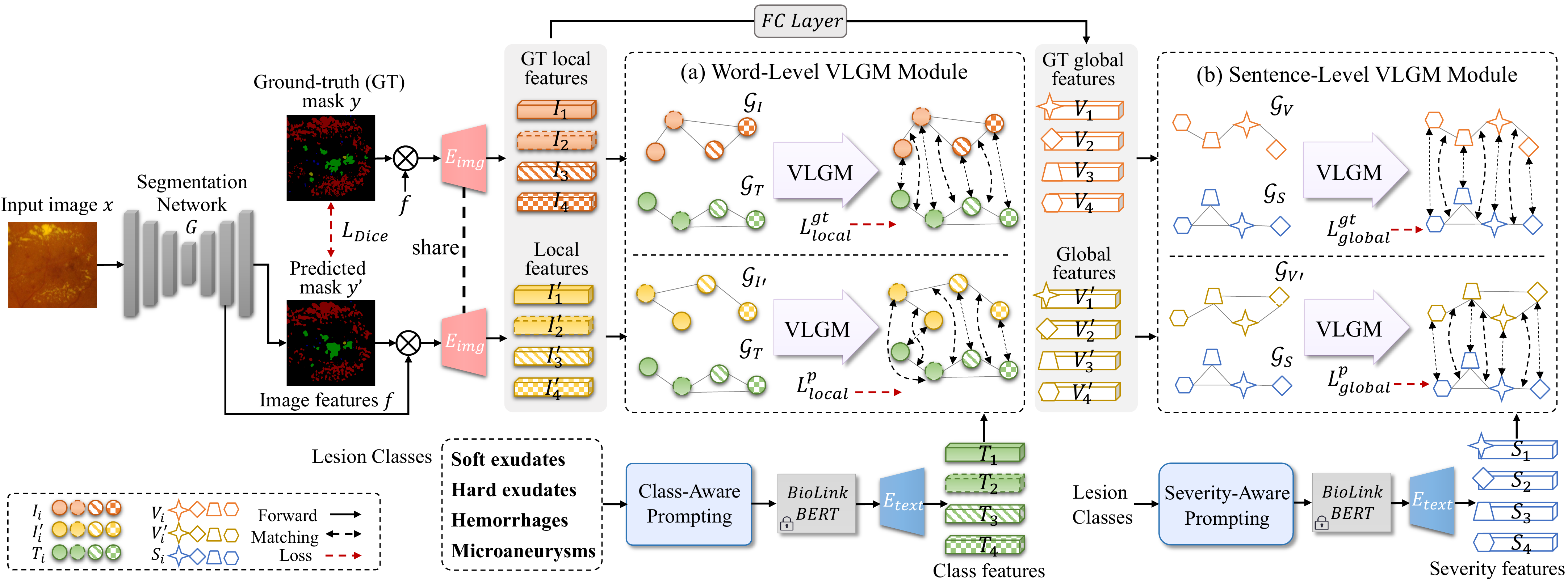}
   \caption{Overview of the proposed Bi-VLGM for text guided medical image segmentation. Our method consists of (a) a world-level VLGM module to match local regions and class-aware prompts, and (b) a sentence-level VLGM module to align the global image and severity-aware prompts.}
   \label{fig:architecture}
\end{figure*}

\section{Related Works}

\subsection{Text Guided Medical Image Segmentation}
Some recent works \cite{wen2022fleak,Tomar2022TGANet,Dai2018Clinical,Li2022LViT, wen2021let,zhou2021cross} utilize the medical reports to facilitate medical image segmentation. Most of these methods \cite{wen2021let,zhou2021cross,wen2022fleak,Tomar2022TGANet,Li2022LViT} apply the medical reports as auxiliary information to assist segmentation. For instance, Li \textit{et al.} \cite{Li2022LViT} introduces a transformer based lesion segmentation model (LViT) to merge the medical text embedding with the image embedding to compensate for the quality deficiency in image data. Without corresponding medical reports, Tomar \textit{et al.} \cite{Tomar2022TGANet} proposes to convert the predicted quantitative information (e.g. number and size of polyps) into text and integrate them into the image features for the final polyps segmentation. Different from the aforementioned methods, Dai \textit{et al.} \cite{Dai2018Clinical} attempts to learn the mapping between medical images and reports to obtain a rough segmentation and further use a multi-sieving convolutional neural network to refine the segmentation results. However, these previous methods directly operate the visual and textual embedding for features fusion or mapping, which ignore the multi-modal heterogeneity gap.  To bridge this gap, our method proposes to perform vision-language graph matching (VLGM) between visual and textual features to measure their semantic difference, in order to make segmentation model selectively learn useful representation to promote segmentation performance.  

\subsection{Medical Vision-Language Models}
Several vision-language models \cite{monajatipoor2022berthop,pan2022vision,chen2022multi,chen2022align,muller2021joint,li2020comparison,boecking2022making,moon2022multi} have been proposed for medical computer vision tasks. Muller \textit{et al.} \cite{muller2021joint} introduces a vision-language model to match the image and text embeddings through the instance-level and local contrastive learning, which can be applied to some downstream tasks. As a result, this method requires a large-scale paired image-text dataset, limiting its practice usage. Recently, BERTHop \cite{monajatipoor2022berthop} is proposed to resolve the data-hungry issue by using a pre-trained BlueBERT \cite{peng2019transfer} and a frequency-aware PixelHop++ \cite{chen2020pixelhop++} to encode text and image to embeddings, respectively. With the appropriate vision and text extractor, BERTHop can efficiently capture the association between the image and text.

Nevertheless, these methods directly align inter-modal features and ignore the relation among the intra-modal features, potentially causing the distortion of the intra-modal feature space. Thus, we consider the intra-modal relation, and introduce a vision-language graph matching (VLGM) to remain the relation consistent by exploiting the high-order relation among visual-textual features.

\section{Method}
As shown in Fig. \ref{fig:architecture}, we propose a Bi-level class-severity-aware Vision-Language Graph Matching (Bi-VLGM) for text guided medical image segmentation, to exploit the relation between local (global) and class (severity) features through vision-language graph matching (VLGM). Given an input image $x$, the segmentation network $G$ outputs the predicted mask $y'$. 
Then, an image encoder $E_{img}$ encodes image features to GT local features $\left \{ I_i \right \}^{C}_{i=1}$ and local features $\left \{ I_i' \right \}^{C}_{i=1}$ with $C$ lesion classes for the ground-truth (GT) and the predicted mask, respectively. Meanwhile, the class-aware prompting and severity-aware prompting take the lesion classes as input and generate medical prompts, which are further fed into a BioLinkBERT \cite{yasunaga2022linkbert} and a text encoder $E_{text}$ to extract the class features  $\left \{ T_i \right \}^{C}_{i=1}$ and the severity features $\left \{ S_i \right \}^{\mathcal{B}}_{i=1}$, respectively.
Afterward, we perform VLGM between local features and class features, and that between the global features $\left \{ V_i' \right \}^{\mathcal{B}}_{i=1}$ integrated with local features and severity features. Similarly, GT local features are matched with class features, and GT global features $\left \{ V_i' \right \}^{\mathcal{B}}_{i=1}$ are aligned with severity features. $E_{img}$ and $E_{text}$ are optimized through $L_{local}^{gt}$ and $L_{global}^{gt}$.
Enabling two encoders frozen, we update $G$ with $L_{local}^p$ and $L_{global}^p$. 


\subsection{Word-Level VLGM Module}

\begin{figure}[t]
  \centering
   \includegraphics[width=0.9\linewidth]{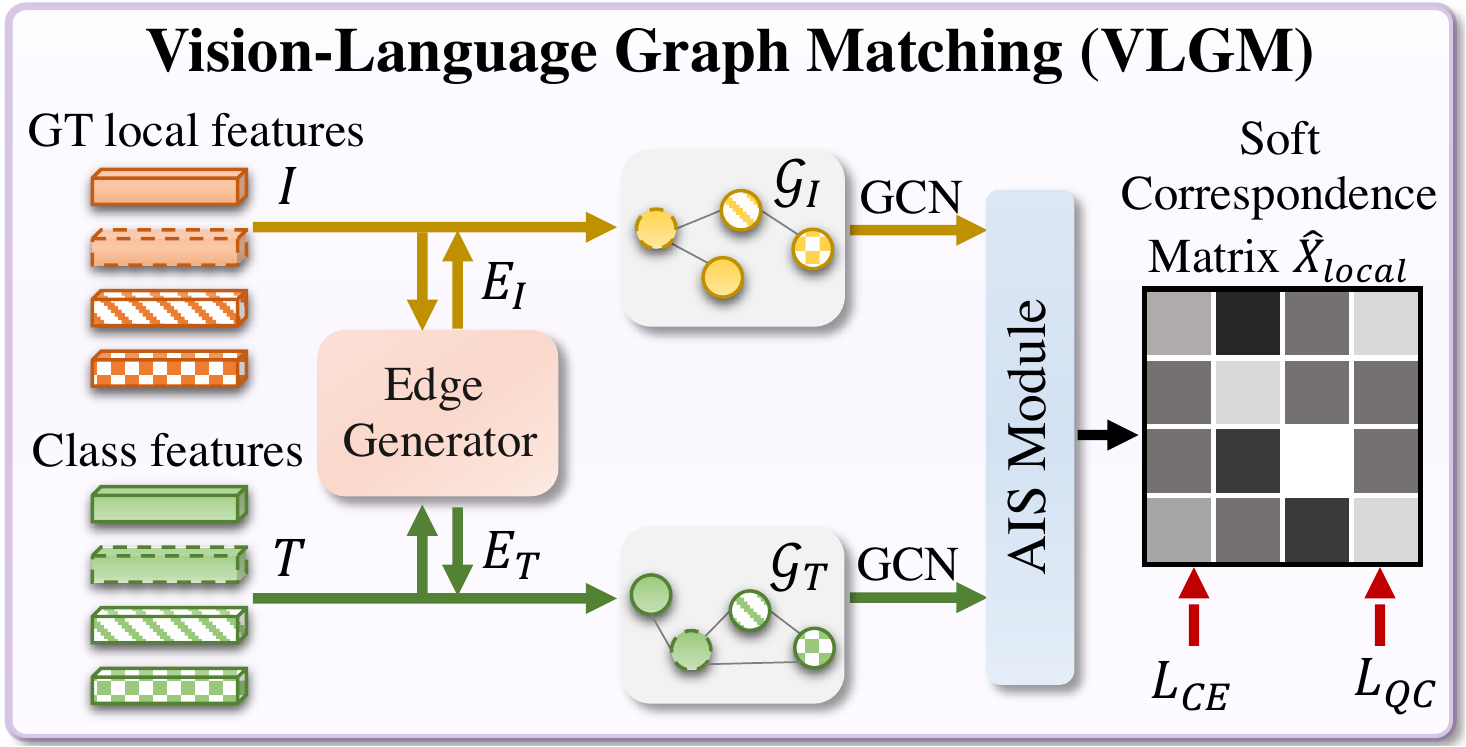}
   \caption{The Vision-Language Graph Matching (VLGM)}
   \label{fig:vlgm}
\end{figure}

To align the local features with class features for each lesion class, we introduce a word-level VLGM module, as shown in Fig. \ref{fig:architecture} (a). This module includes a local-class alignment for GT masks and a local-class alignment for predicted masks, where the former aims to bridge the gap among GT local-class features, and the latter measures the semantic difference between local and class features to make local features to preserve more class-aware context for segmentation. Moreover, to remain the intra-modal relation during alignment, we reformulate the vision-language matching as graph matching \cite{gao2021deep, fu2021robust} problem and introduce a vision-language graph matching (VLGM) to implement the alignment, which first constructs graph for each feature and then perform graph matching, as depicted in Fig. \ref{fig:vlgm}.



\paragraph{Local-Class Alignment for GT Masks.} 
To encourage the image and text encoder to capture better semantic representation for local regions and class-aware prompts, we perform VLGM between GT local features $I$ for GT mask and class features $T$ for class-aware prompts. Given a GT mask $y \in \mathbb{R}^{C\times H \times W}$ and the upsampled image features $f \in \mathbb{R}^{D \times H \times W}$ from segmentation network, we first compute matrix multiplication between $y$ and $f$, and then encode the output with an image encoder $E_{img}$ to obtain the GT local features $\left \{ I_i \right \}^{C}_{i=1} \in \mathbb{R}^{C\times D}$ for $C$ lesion classes, where $D$ denotes the feature dimension, $H$ and $W$ represent the height and width of input image.
As for class-aware prompts, we introduce a \textit{\textbf{Class-Aware Prompting}} to generate medical prompts for each lesion class, e.g. ``\texttt{A fundus image with [CLS]}". \texttt{[CLS]} indicates the name of lesion class. With this prompt engineering, we generate the class-aware prompts for all the lesion classes, and feed them into a pre-trained BioLinkBERT \cite{yasunaga2022linkbert} and a text encoder $E_{text}$ sequentially to extract the class features $\left \{ T_i \right \}^{C}_{i=1} \in \mathbb{R}^{C\times D}$ for $C$ lesion classes.


As illustrated in Fig. \ref{fig:vlgm}, we regard the GT local features $I = \left \{ I_i \right \}^{C}_{i=1}$ and class features $T=\left \{ T_i \right \}^{C}_{i=1}$ as initial graph node features, and build their graphs $\mathcal{G}_I=\left \{I,E_I \right \}$ and $\mathcal{G}_T=\left \{T,E_T\right \}$, respectively. 
To obtain the graph edges $E_I$ and $E_T$, we apply an edge generator \cite{fu2021robust} to the graph node features $I$ and $T$. Specifically, the edge generator first uses a transformer to learn the soft edges of any two nodes in the graph, and then adopts a softmax function on the inner product of the soft edge features to acquire the soft edge adjacency matrices $E_I$, $E_T$.
These graph edges reveal the high-order relation for GT local features and class features. To capture the high-order relation, we embed both the graph nodes (local/class features) and high-order graph structure (edges) into node feature space through the graph convolutional networks (GCN) \cite{fu2021robust} to obtain new node features $GCN(E_I, I)$ for GT local features and $GCN(E_T, T)$ for class features.

With graphs $\mathcal{G}_I$, $\mathcal{G}_T$ for GT local features and class features, we perform VLGM between them to narrow their gap in graphic space. We employ the AIS module \cite{fu2021robust} to predict the soft correspondence matrix $\hat{X}_{local}=AIS(GCN(E_I, I),GCN(E_T, T))$, which represents the possibility of establishing a matching relation between any pair of nodes in two graphs. 
The AIS module \cite{fu2021robust} consists of an affinity layer to compute an affinity matrix between two graphs, instance normalization to make the element of the affinity matrix positive, and Sinkhorn\cite{sinkhorn1964relationship} to handle outliers in the affinity matrix. 
Then, to supervise the prediction of soft correspondence matrix $\hat{X}_{local}$, we compute the cross-entropy loss between the ground-truth correspondence matrix $X_{local}^{gt}$ and $\hat{X}_{local}$, which is defined as:
\begin{equation}
\begin{aligned}
    L_{CE} (\hat{X}_{local}, & X_{local}^{gt})  = - \sum_{i}^{N}\sum_{j}^{M} (X_{local}^{gt}(i,j)log\hat{X}_{local}(i,j)  \\ 
& + (1-X_{local}^{gt}(i,j))log(1-\hat{X}_{local}(i,j)),
\end{aligned}
\end{equation}
where $N$ and $M$ indicate the number of row and column of $\hat{X}_{local}$. 
Each element in $\hat{X}_{local}\in \mathbb{R}^{N\times M}$ and $X_{local}^{gt} \in \mathbb{R}^{N\times M}$ indicates the high-order relation between local features and class features. The element of $X_{local}^{gt}$ is set $1$, where the lesion class for the local features is highly correlated to the lesion name for the class features. By optimizing $L_{CE}$, we can make the relation of local-class features get close to that of GT local-class features, in order to align the local features with class features.

For the consistency of the high-order relation of $\mathcal{G}_I$ and $\mathcal{G}_T$, we use the Quadratic Constrains (QC) \cite{li2022sigma} to minimize their structural discrepancy, which is formulated as:
\begin{equation}
L_{QC} (E_I, E_T, \hat{X}_{local}) = \frac{1}{N M} \sum_{i}^{N}\sum_{j}^{M} (E_I \hat{X}_{local} - \hat{X}_{local} E_T)_{i,j}
\end{equation}
The overall local-class loss for GT masks is defined as:
\begin{equation}
    L_{local}^{gt} = L_{CE}(\hat{X}_{local}, X_{local}^{gt}) + L_{QC}(E_I,E_T,\hat{X}_{local}),
\end{equation}

With $ L_{local}^{gt}$, we can ensure the representational capacity of image and text encoders at the initial training stage, which can reduce the ambiguities brought by using local features for the predicted mask that is of low confidence. 

\paragraph{Local-Class Alignment for Predicted Masks.}
With the effective encoders for semantic feature extraction, we leverage VLGM to measure the semantic distance between local features for predicted mask $y'$ and class features for class-aware prompts. Given the predicted mask $y'$ and image features $f$, we first perform the matrix multiplication of $y'$ and $f$, and then feed the output into image encoder $E_{img}$ to obtain local features $I'=\left \{ {I'}_i \right \}^{C}_{i=1} \in \mathbb{R}^{C\times D}$ for $C$ lesion classes.
Afterwards, we adopt $I'$ as initial graph node features and construct a graph $\mathcal{G}_{I'}=\left \{I',E_{I'} \right \}$, where $E_{I'}$ represents edges generated by an edge generator \cite{fu2021robust} with node features $I'$. To grasp the high-order relation, we embed the graph edges into node feature space through $GCN$ and obtain the new node features $GCN(E_{I'}, I')$ for $\mathcal{G}_{I'}$.
With graphs $\mathcal{G}_{I'}$, $\mathcal{G}_T$ of local features and class features, we first utilize the AIS module \cite{fu2021robust} to produce the soft correspondence matrix $\tilde{X}_{local} = AIS(GCN(E_{I'}, I'), GCN(E_T, T))$ and then compute the local-class loss for predicted masks $L_{local}^p$ to measure their semantic difference, which is formulated as:
\begin{equation}
    L_{local}^p = L_{CE}(\tilde{X}_{local}, X_{local}^{gt}) + L_{QC}(E_{I'},E_T,\tilde{X}_{local}),
\end{equation}
where $ X_{local}^{gt}$ indicates GT correspondence matrix. 
By adopting the local-class loss for predicted masks $L_{local}^p$, the segmentation model can capture more class-aware information in local features that cover each class region, making the predicted mask more accurate.

With word-level VLGM module, we can bridge the gap between the local features for each class region and class features for each lesion class by minimizing their semantic difference in graphical space, and further equip the segmentation model with more class-aware representation to promote segmentation performance. 

\subsection{Sentence-Level VLGM Module} 
\noindent As illustrated in Fig. \ref{fig:architecture} (b), we introduce a sentence-level VLGM module to include severity information and align global image features with the severity-aware prompts. This module consists of a global-severity alignment for GT masks to bridge the gap between global and severity features, and a global-severity alignment for predicted masks to equip the segmentation model with more severity information to boost segmentation performance. 

\paragraph{Global-Severity Alignment for GT Masks.}
To force the encoders to capture more semantic information for global images and severity-aware prompts, we align GT global features $I'$ for global images with severity features $S$ for severity-aware prompts. Given GT local features $\left \{ I_i \right \}^{\mathcal{B}}_{i=1} \in \mathbb{R}^{\mathcal{B} \times C \times D}$ from word-level VLGM module, we map them to GT global features $\left \{ V_i \right \}^{\mathcal{B}}_{i=1} \in \mathbb{R}^{\mathcal{B} \times 1 \times D}$ by a fully connected layer and computing mean along channel dimension, where $\mathcal{B}$, $C$ and $D$ denote the batch size, number of lesion classes and feature dimension, respectively.  As for severity-aware prompts, we introduce a \textit{\textbf{Severity-Aware Prompting}} to automatically generates medical prompts that quantify the severity level of retinal lesion in global view. As listed in Table \ref{tab:prompts}, we design five templates with reference to previous report generation methods \cite{wu2017generative,mishra2020automatic} to make medical prompts diverse. Each medical prompt includes the lesion name \texttt{[CLS]} and its adjective \texttt{[ADJ]}, where the adjective quantifies the severity level based on lesion ratio. The lesion ratio is area ratio of each lesion class. We set two thresholds $t_1$,$t_2$ to divide lesion ratio into three levels. There are three groups of adjectives to describe three levels, i.e. \texttt{few/some/many}, \texttt{low/medium/high-density}, and \texttt{low/medium/high-severity}. In addition, we use `\texttt{and}' to connect a second lesion class when there are more than one lesion class in the input image. 
An example for a retinal image with few hard exudates and many hemorrhages is provided,
``\texttt{This fundus image has low-density hard exudates and high-severity hemorrhages.}". 
With severity-aware prompts, we feed them into a BioLinkBERT and text encoder sequentially to obtain severity features $\left \{ S_i \right \}^{\mathcal{B}}_{i=1}$ 

\begin{table}[t]
	\centering
	\caption{The severity-aware prompting. }
	\scalebox{0.9}{
 \begin{tabular}{cccc}
			\toprule[1pt]
			\multicolumn{4}{c}{Medical Prompts} \\ \hline
\multicolumn{4}{l}{1. This fundus image has [ADJ] [CLS].} \\
\multicolumn{4}{l}{2. There are [ADJ] [CLS] in this fundus image.}  \\
\multicolumn{4}{l}{3. A fundus image with [ADJ] [CLS].}  \\
\multicolumn{4}{l}{4. A diabetic retinopathy image has [ADJ] [CLS].} \\
\multicolumn{4}{l}{5. [ADJ] [CLS] in a diabetic retinopathy fundus image.} \\ 
			\bottomrule[1pt]
	\end{tabular}
	}
	\label{tab:prompts}
\end{table}

Given the GT global features $\left \{ V_i \right \}^{\mathcal{B}}_{i=1}$  and severity features $\left \{ S_i \right \}^{\mathcal{B}}_{i=1}$, we regard these features as initial graph node features and construct their graphs $\mathcal{G}_V=\left \{V,E_V \right \}$, $\mathcal{G}_S=\left \{S,E_S\right \}$, where $E_V$ and $E_S$ represent their edges generated by edge generator \cite{fu2021robust}. To leverage the high-order relation, we embed the edge features into node features by computing their new node features $GCN(E_V, V)$ and $GCN(E_S, S)$.
\begin{table*}[t]
	\centering
	\caption{Quantitative comparison on IDRiD dataset.}
	\scalebox{0.80}{
		\begin{tabular}{p{3.8cm}<{\centering}|p{0.7cm}<{\centering}p{0.7cm}<{\centering}p{0.7cm}<{\centering}p{0.7cm}<{\centering}|p{1.05cm}<{\centering}|p{0.7cm}<{\centering}p{0.7cm}<{\centering}p{0.7cm}<{\centering}p{0.7cm}<{\centering}|p{0.7cm}<{\centering}|p{0.7cm}<{\centering}p{0.7cm}<{\centering}p{0.7cm}<{\centering}p{0.7cm}<{\centering}|p{0.7cm}<{\centering}} 
			\toprule[1pt]
			\multirow{2}{*}{Methods} & \multicolumn{5}{c|}{AUPR} & \multicolumn{5}{c|}{F} & \multicolumn{5}{c}{IoU} \\ \cline{2-16}
			 &  EX & HE & SE & MA & mAUPR & EX & HE & SE & MA & mF & EX & HE & SE & MA & mIoU \\ \hline
            DNL \cite{yin2020disentangled} (2020) & 75.12 & 64.04 & 64.73 & 32.48 & 59.09 & 73.15 & 61.87 & 63.96 & 32.78 & 57.94 & 57.67 & 44.80 & 47.03 & 19.61 & 42.28\\
            SPNet \cite{hou2020strip} (2020)& 77.22 & 68.13 & 70.90 & 41.90 & 64.54 & 75.16 & 66.53 & 68.96 & 42.45 & 63.27 & 60.21 & 49.85 & 52.62 & 26.94 & 47.40\\
            HRNetV2 \cite{wang2020deep} (2020) & 80.38 & 65.25 & 68.67 & 44.25 & 64.64 & 78.35 & 63.77 & 67.58 & 44.63 & 63.58 & 64.41 & 46.81 & 51.04 & 28.72 & 47.74 \\
            Swin-B \cite{liu2021swin} (2021) & 81.30 & 67.70 & 66.46 & 44.19 & 64.91 & 79.64 & 66.42 & 66.00 & 44.09 & 64.04 & 66.17 & 49.72 & 49.27 & 28.28 & 48.36 \\ 
            Twins-SVT-B \cite{chu2021twins} (2021) & 80.09 & 63.12 & 68.86 & 43.27 & 63.84 & 78.56 & 61.98 & 68.19 & 42.42 & 62.79 & 64.68 & 44.91 & 51.76 & 26.92 & 47.07 \\
            M2MRF \cite{liu2021m2mrf} (2021) & 82.16 & 68.69 & 69.32 & \textbf{48.80} & \underline{67.24} & \underline{79.85} & 66.42 & 67.92 & \textbf{48.63} & 65.71 & 66.46 & 49.72 & 51.43 & \textbf{32.13} & 49.9 \\ 
            PCAA \cite{Liu_2022_CVPR} (2022) & 81.63 & 66.74 & \textbf{75.49} & 43.33 & 66.80 & 79.58 & 64.59 & \textbf{74.13} & 43.17 & 65.37 & 66.09 & 47.70 & \textbf{58.89} & 27.53 & 50.05 \\ 
            IFA \cite{hu2022learning} (2022) & 81.92 & \underline{69.01} & 70.47 & \underline{46.35} & 66.94 & 79.80 & \textbf{67.43} & 69.12 &  \underline{46.35} & \underline{65.68} & 66.39 & \textbf{50.86} & 52.82 & \underline{30.17} & \underline{50.06} \\
            LViT \cite{Li2022LViT} (2022) & \underline{82.19} & 63.36 & 70.32 & 43.65 & 64.88 & 79.99 & 60.96 & 69.33 & 43.44 & 63.43 & 66.65 & 43.85 & 53.06 & 27.74 & 47.82 \\
            TGANet \cite{Tomar2022TGANet} (2022) & 82.16 & 65.60 & 68.86 & 42.19 & 64.70 & 80.01 & 63.46 & 67.89 & 41.29 & 63.16 & \underline{66.67} & 46.48 & 51.39 & 26.01 & 47.64 \\ 
            \hline
Bi-VLGM & \textbf{82.48} & \textbf{69.32} & \underline{74.50} & 46.20 & \textbf{68.12} & \textbf{80.51} & \underline{67.42} & \underline{72.95} & 45.98 & \textbf{66.71} & \textbf{67.38} & \underline{50.85} & \underline{57.41} & 29.85 & \textbf{51.37} \\ 
			\bottomrule[1pt]
	\end{tabular}
	}
	\label{sota_idrid}
\end{table*}
With graphs $\mathcal{G}_V$, $\mathcal{G}_S$ for GT global features and severity features, we compute their soft correspondence matrix $\hat{X}_{global} = AIS(GCN(E_V, V), GCN(E_S, S))$ through AIS module \cite{fu2021robust}, and calculate the global-severity loss for GT masks $L_{global}^{gt}$, which is formulated as:
\begin{equation}
    L_{global}^{gt} = L_{CE}(\hat{X}_{global}, X_{global}^{gt}) + L_{QC}(E_V,E_S,\hat{X}_{global}),
\end{equation}
where $X_{global}^{gt}$ represents the GT correspondence matrix. Each element of $X^{gt}_{global}$ indicates the relation between any pair of global-severity features within batch, where the relation is set as $1$ when the severity-aware prompts for severity features can reflect the severity level of the input image for the global features.
With the supervision of global-severity loss for GT masks, the encoders are more capable of extracting semantic information from global image and severity-aware prompts.


\paragraph{Global-Severity Alignment for Predicted Masks.} 
With effective encoders for capturing semantic representation, we can leverage them to measure the semantic difference between global and severity features, to guide the learning of the predicted mask. The global features $\left \{ V_i' \right \}^{\mathcal{B}}_{i=1} \in \mathbb{R}^{\mathcal{B}\times 1 \times D}$ are projected from local features $\left \{ I_i' \right \}^{\mathcal{B}}_{i=1} \in \mathbb{R}^{\mathcal{B} \times C \times D} $ by a fully connected layer and computing mean along channel dimension.
Given the global features $\left \{ V_i' \right \}^{\mathcal{B}}_{i=1}$, we use these features as initial graph node features for graph $\mathcal{G}_{V'}=\left \{V',E_{V'} \right \}$, where $E_{V'}$ represents edges generated by edge generator \cite{fu2021robust}, and employ $GCN$ to obtain the new node features $GCN(E_{V'}, V')$. 
With graphs $\mathcal{G}_{V'}$, $\mathcal{G}_{S}$ of global features  and severity features, we first compute their soft correspondence matrix $\tilde{X}_{global} = AIS(GCN(E_{V'}, V'), GCN(E_S, S))$ through AIS module \cite{fu2021robust}, and then calculate the global-severity loss for predicted masks $L_{global}^{p}$, which is defined as:
\begin{equation}
    L_{global}^{p} = L_{CE}(\tilde{X}_{global}, X_{global}^{gt}) + L_{QC}(E_{V'},E_S,\tilde{X}_{global}),
\end{equation}
By minimizing $L_{global}^{p}$, the segmentation model can equip the global features with more severity features to improve segmentation performance.


\subsection{Overall Loss Functions}
\noindent Eventually, to optimize the image and text encoders, the overall loss function is formulated as:
\begin{equation}
    L_E = \lambda_a L_{local}^{gt} + \lambda_b L_{global}^{gt},
\end{equation}
where $\lambda_a$ and $\lambda_b$ are the weighting coefficients to balance the influence of each loss. 
At the initial training stage, the image and text encoders are not able to extract semantic information, which may cause the ambiguities brought by using the visual features for the predicted mask that is of low confidence. $L_E$ aims to avoid the ambiguities before the optimization of the segmentation network.

To optimize the segmentation network, we utilize the Dice loss $L_{Dice}$ \cite{milletari2016v}, $L_{local}^p$, and $L_{global}^p$. The overall loss function is defined as:
\begin{equation}
    L_G = \lambda_c L_{Dice} + \lambda_d L_{local}^p  + \lambda_e L_{global}^p, 
\end{equation}
where $\lambda_c$, $\lambda_d$ and $\lambda_e$ denote the hyper-parameters to weight the three losses.
With $L_E$ and $L_G$, we can bridge the gap between local (global) and class (severity) features and make the segmentation model selectively learn the class-severity-aware information to promote segmentation.

    


\begin{table*}[t]
	\centering
	\caption{Quantitative comparison on DDR dataset.}
	\scalebox{0.80}{
		\begin{tabular}{p{3.8cm}<{\centering}|p{0.7cm}<{\centering}p{0.7cm}<{\centering}p{0.7cm}<{\centering}p{0.7cm}<{\centering}|p{1.05cm}<{\centering}|p{0.7cm}<{\centering}p{0.7cm}<{\centering}p{0.7cm}<{\centering}p{0.7cm}<{\centering}|p{0.7cm}<{\centering}|p{0.7cm}<{\centering}p{0.7cm}<{\centering}p{0.7cm}<{\centering}p{0.7cm}<{\centering}|p{0.7cm}<{\centering}} 
			\toprule[1pt]
			\multirow{2}{*}{Methods} & \multicolumn{5}{c|}{AUPR} & \multicolumn{5}{c|}{F} & \multicolumn{5}{c}{IoU} \\ \cline{2-16}
			 &  EX & HE & SE & MA & mAUPR & EX & HE & SE & MA & mF & EX & HE & SE & MA & mIoU \\ \hline
            DNL \cite{yin2020disentangled} (2020) & 56.05 & 47.81 & 42.01 & 14.71 & 40.14 & 53.36 & 42.71 & 40.40 & 15.60 & 38.02 & 36.39 & 27.15 & 25.33 & 8.46 & 24.33\\
            SPNet \cite{hou2020strip} (2020)& 44.10 & 38.22 & 32.93 & 12.37 & 31.91 & 38.78 & 24.13 & 34.00 & 13.74 & 27.66 & 24.19 & 13.76 & 20.55 & 7.38 & 16.47\\
            HRNetV2 \cite{wang2020deep} (2020) & 61.55 & 45.68 & 46.91 & 26.70 & 45.21 & 58.98 & 44.96 & 44.86 & 26.99 & 43.95 & 41.82 & 29.01 & 28.94 & 15.60 & 28.84 \\
            Swin-B \cite{liu2021swin} (2021) & \underline{62.95} & 53.46 & 50.56 & 23.46 & 47.61 & \underline{60.12} & 51.10 & \underline{50.85} & 23.38 & \underline{46.36} & \underline{42.98} & 34.42 & \underline{34.15} & 13.24 & \underline{31.20} \\ 
            Twins-SVT-B \cite{chu2021twins} (2021) & 59.71 & 49.96 & \textbf{52.72} & 22.03 & 46.11 & 56.83 & 45.04 & \textbf{53.19} & 21.54 & 44.15 & 39.70 & 29.08 & \textbf{36.24} & 12.07 & 29.28 \\
            M2MRF \cite{liu2021m2mrf} (2021) & \textbf{63.59} & 54.43 & 49.35 & \textbf{28.38} & \underline{48.94} & \textbf{60.62} & 45.16 & 47.78 & \textbf{28.04} & 45.40 & \textbf{43.49} & 29.17 & 31.39 & \textbf{16.31} & 30.09 \\ 
            PCAA \cite{Liu_2022_CVPR} (2022) & 60.57 & \textbf{57.46} & 41.49 & 18.58 & 44.52 & 56.89 & \textbf{54.47} & 36.68 & 20.57 & 42.15 & 39.76 & \textbf{37.43} & 22.46 & 11.47 & 27.78 \\ 
            IFA \cite{hu2022learning} (2022) & 61.51 & 46.19 & 48.90 & 12.98 & 42.40 & 56.76 & 46.28 & 48.25 &  0.55 & 37.96 & 39.62 & 30.11 & 31.80 &  0.28 & 25.45 \\
            LViT \cite{Li2022LViT} (2022) & 61.35 & 46.29 & 48.06 & \underline{27.61} & 45.83 & 59.15 & 42.85 & 46.88 & \underline{27.78} & 44.17 & 42.00 & 27.27 & 30.62 & \underline{16.13} & 29.00 \\
            TGANet \cite{Tomar2022TGANet} (2022) & 60.49 & 52.63 & 43.55 & 26.81 & 45.87 & 58.92 & 42.19 & 41.27 & 26.92 & 42.32 & 41.76 & 26.73 & 26.00 & 15.55 & 27.51 \\
            \hline 
            Bi-VLGM  & 62.01 & \underline{57.38} & \underline{50.95} & 26.19 & \textbf{49.13} & 57.90 & \underline{54.38} & 50.81 & 26.06 & \textbf{47.29} & 40.75 & \underline{37.34} & 34.06 & 14.98 & \textbf{31.78} \\

			\bottomrule[1pt]
	\end{tabular}
	}
	\label{sota_ddr}
\end{table*}

\section{Experiments}
\label{sec:experiments}
\subsection{Training Details}

\paragraph{Datasets.} We conduct the experiments on two main publicly available datasets, i.e. the \textbf{IDRiD} \cite{porwal2020idrid} and \textbf{DDR} \cite{li2019diagnostic} datasets. The \textbf{IDRiD} dataset contains 81 colour fundus images with the resolution of $4288 \times 2848$, among which 54 for training and 27 for testing. The \textbf{DDR} dataset consists of 757 colour fundus images with size ranging from $1088 \times 1920$ to $3456 \times 5184$. It contains 383 images for training, 149 images for validation and the rest 225 for testing. These two datasets are provided with pixel-level annotation for four different retinal lesions, i.e. soft exudates (SE), hard exudates (EX), microaneurysms (MA) and hemorrhages (HE). We resize images into $1024\times 1024$ and $1440 \times 960$ for DDR and IDRiD datasets, respectively. 

\paragraph{Implementation Details.} We adopt the HRNetV2 \cite{wang2020deep} as our segmentation model. A pre-trained BioLinkBERT \cite{yasunaga2022linkbert} is utilized to encode medical prompts to features. During the training process, we alternate training between the segmentation network and the image and the text encoders with $L_G$ and $L_E$, respectively. In $L_{local}^p$ and $L_{global}^p$, we replace the cross entropy loss with L1 loss to provide a smooth constraint to the segmentation network. Our method is implemented with the PyTorch library on NVIDIA A100. SGD optimizer is adopted to optimize the model parameters for a maximum of 40,000 iterations for the IDRiD dataset and 60,000 for the DDR dataset, respectively. The initial learning rate is 0.01 with the poly policy. The batch size is 8. Hyper-parameters $\lambda_a, \lambda_b, \lambda_c, \lambda_d$, $\lambda_e$, $t_1$ and $t_2$ are set as 0.5, 0.5, 1.0, 0.5, 0.5, 0.06 and 0.12, respectively. For evaluation metrics, we utilize the commonly used Area Under Precision-Recall curve (AUPR), F-score, intersection-over-union (IoU), and their mean values (mAUPR, mF and mIoU).
\subsection{Results on the IDRiD dataset}
\noindent We first verify the effectiveness of our method on the IDRiD dataset, and the corresponding comparison results are listed in Table \ref{sota_idrid} with the first and second best results highlighted in bold and underline. We compare our method with the text guided medical image segmentation methods \cite{Li2022LViT, Tomar2022TGANet} and the state-of-the-art segmentation methods \cite{liu2021m2mrf, yin2020disentangled,hou2020strip, wang2020deep,liu2021swin,chu2021twins,Liu_2022_CVPR,hu2022learning} for four lesion classes, i.e. EX, HE, SE and MA. The proposed Bi-VLGM achieves the best segmentation performance with the mAUPR, mF and mIoU scores of 68.12\%, 66.71\% and 51.37\%, surpassing the existing methods by a large margin. In terms of performance for each lesion class, our method arrives at the best and second best on 3 out of 4 categories in AUPR score, implying our superior performance. 

\begin{figure}[t]
  \centering
   \includegraphics[width=\linewidth]{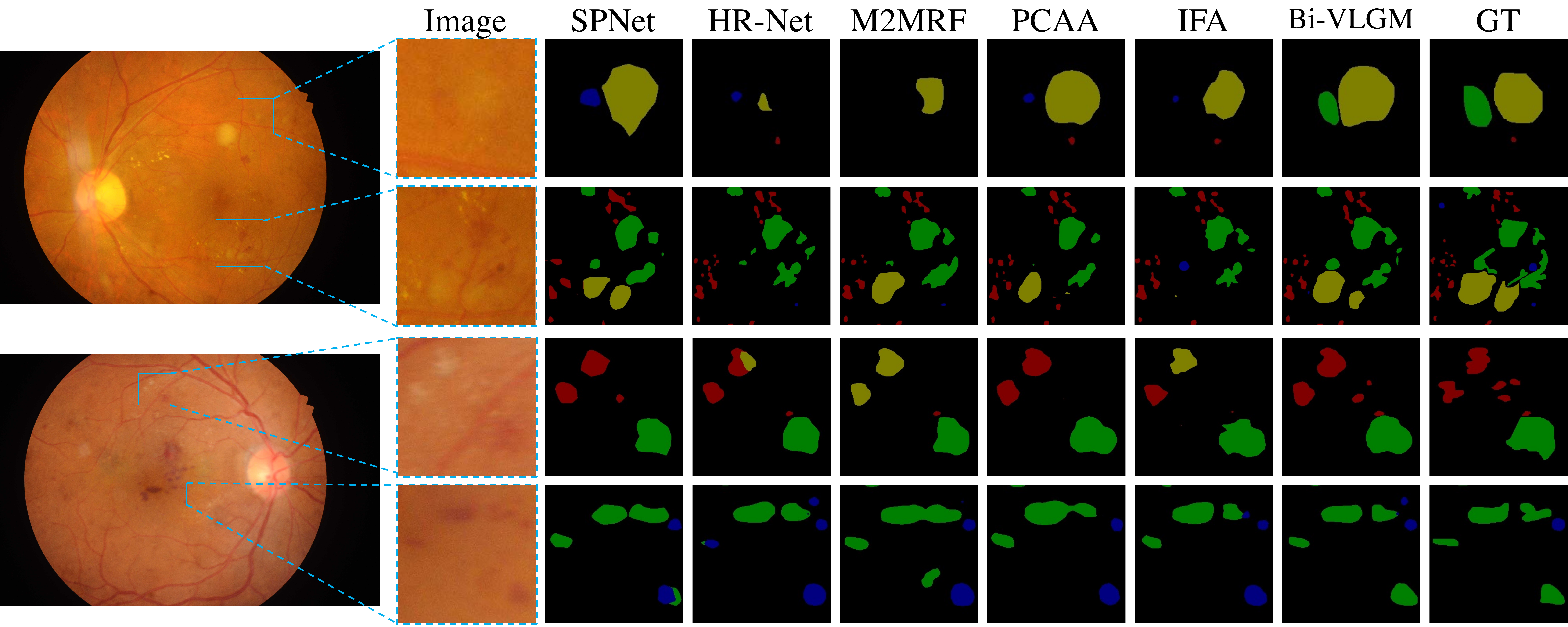}
   \caption{The visual results on the IDRiD dataset in comparison with the state-of-the-art methods. The regions filled in red, green, yellow and blue are EX, HE, SE and MA, respectively.} 
   \label{fig:vis_idrid}
\end{figure}

\begin{table*}[t]
	\centering
	\caption{Ablation studies on IDRiD dataset.}
 
	\scalebox{0.85}{
		\begin{tabular}{p{1.2cm}<{\centering}|p{1.2cm}<{\centering}|p{0.7cm}<{\centering}p{0.7cm}<{\centering}p{0.7cm}<{\centering}p{0.7cm}<{\centering}|p{1.05cm}<{\centering}|p{0.7cm}<{\centering}p{0.7cm}<{\centering}p{0.7cm}<{\centering}p{0.7cm}<{\centering}|p{0.7cm}<{\centering}|p{0.7cm}<{\centering}p{0.7cm}<{\centering}p{0.7cm}<{\centering}p{0.7cm}<{\centering}|p{0.7cm}<{\centering}} 
			\toprule[1pt]
			&  & \multicolumn{5}{c|}{AUPR} & \multicolumn{5}{c|}{F} & \multicolumn{5}{c}{IoU} \\ \hline 
			Word & Sentence &  EX & HE & SE & MA & mAUPR & EX & HE & SE & MA & mF & EX & HE & SE & MA & mIoU \\ \hline
            &  & 80.38 & 65.25 & 68.67 & 44.25 & 64.64 & 78.35 & 63.77 & 67.58 & 44.63 & 63.58 & 64.41 & 46.81 & 51.04 & 28.72 & 47.74 \\
            \checkmark & & 80.44 & 66.30 & \textbf{76.31} & 42.78 & 66.46 & 78.44 & 65.10 & \textbf{74.63} & 43.33 & 65.38 & 64.53 & 48.26 & \textbf{59.53} & 27.66 & 50.00 \\
             & \checkmark & 82.25 & 68.30 & 73.82 & 45.01 & 67.35 & 80.39 & 66.66 & 72.63 & 43.47 & 65.78 & 67.20 & 49.99 & 57.02 & 27.77 & 50.50 \\
            \checkmark & \checkmark & \textbf{82.48} & \textbf{69.32} & 74.50 & \textbf{46.20} & \textbf{68.12} & \textbf{80.51} & \textbf{67.42} & 72.95 & \textbf{45.98} & \textbf{66.71} & \textbf{67.38} & \textbf{50.85} & 57.41 & \textbf{29.85} & \textbf{51.37} \\ 
			\bottomrule[1pt]
	\end{tabular}
	}
	\label{ablation}
\end{table*}

\begin{table*}[t]
	\centering
	\caption{The effectiveness of the severity-aware prompting with different numbers of medical prompts.}
	\scalebox{0.85}{
		\begin{tabular}{p{1.2cm}<{\centering}|p{0.7cm}<{\centering}p{0.7cm}<{\centering}p{0.7cm}<{\centering}p{0.7cm}<{\centering}|p{1.05cm}<{\centering}|p{0.7cm}<{\centering}p{0.7cm}<{\centering}p{0.7cm}<{\centering}p{0.7cm}<{\centering}|p{0.7cm}<{\centering}|p{0.7cm}<{\centering}p{0.7cm}<{\centering}p{0.7cm}<{\centering}p{0.7cm}<{\centering}|p{0.7cm}<{\centering}} 
			\toprule[1pt]
			& \multicolumn{5}{c|}{AUPR} & \multicolumn{5}{c|}{F} & \multicolumn{5}{c}{IoU} \\ \hline 
			\#prompts &  EX & HE & SE & MA & mAUPR & EX & HE & SE & MA & mF & EX & HE & SE & MA & mIoU \\ \hline
1 & 82.45 & 68.28 & 71.74 & 46.46 & 67.24 & 80.43 & 66.97 & 70.50 & 46.15 & 66.01 & 67.27 & 50.34 & 54.44 & 30.00 & 50.51 \\
2 & 82.09 & 69.20 & 73.05 & 47.31 & 67.91 & 80.26 & 67.22 & 71.89 & 45.94 & 66.33 & 67.03 & 50.63 & 56.12 & 29.82 & 50.90 \\
3 & \textbf{82.56} & 67.59 & 72.95 & \textbf{47.55} & 67.66 & \textbf{80.70} & 65.12 & 72.04 & \textbf{47.61} & 66.37 & \textbf{67.64} & 48.28 & 56.30 & \textbf{31.24} & 50.87 \\
4 & 81.84 & 68.17 & 73.48 & 46.92 & 67.60 & 79.92 & 66.50 & 72.59 & 46.77 & 66.44 & 66.56 & 49.82 & 56.97 & 30.52 & 50.97 \\
5 & 82.48 & \textbf{69.32} & \textbf{74.50} & 46.20 & \textbf{68.12} & 80.51 & \textbf{67.42} & \textbf{72.95} & 45.98 & \textbf{66.71} & 67.38 & \textbf{50.85} & \textbf{57.41} & 29.85 & \textbf{51.37} \\ 
			\bottomrule[1pt]
	\end{tabular}
	}
	\label{prompt}
\end{table*}

In Fig. \ref{fig:vis_idrid}, we visualize the segmented results on the IDRiD dataset, and compare our method against existing methods.
In the first row, it is observed that existing methods cannot recognize HE regions while our method can segment out most of HE pixels, indicating that our method is more capable of extracting the features of HE. As shown in the third row, there are some mistakes between EX and SE segmentation in HR-Net \cite{wang2020deep}, M2MRF \cite{liu2021m2mrf} and IFA \cite{hu2022learning}, since EX and SE have similar appearance with yellow or white color, making these two classes misclassified. In contrast, our Bi-VLGM recognizes well on these similar classes, implying that Bi-VLGM can distinguish the tiny difference between the features of two similar classes. 

\subsection{Results on the DDR dataset}
\noindent We further evaluate the proposed Bi-VLGM on the DDR dataset, and the quantitative comparison results are shown in Table \ref{sota_ddr}. In comparison with the text guided medical image segmentation, Bi-VLGM surpasses the LViT \cite{Li2022LViT} and TGANet \cite{Tomar2022TGANet} by more than 3\% of mAUPR. When comparing with the state-of-the-art segmentation methods, the performance of Bi-VLGM outperforms that of M2MRF \cite{liu2021m2mrf} and PCAA \cite{Liu_2022_CVPR} by a large margin of 1.69\% and 4.0\% in mIoU, indicating our superiority to existing methods.

\begin{figure}[t]
  \centering
   \includegraphics[width=\linewidth]{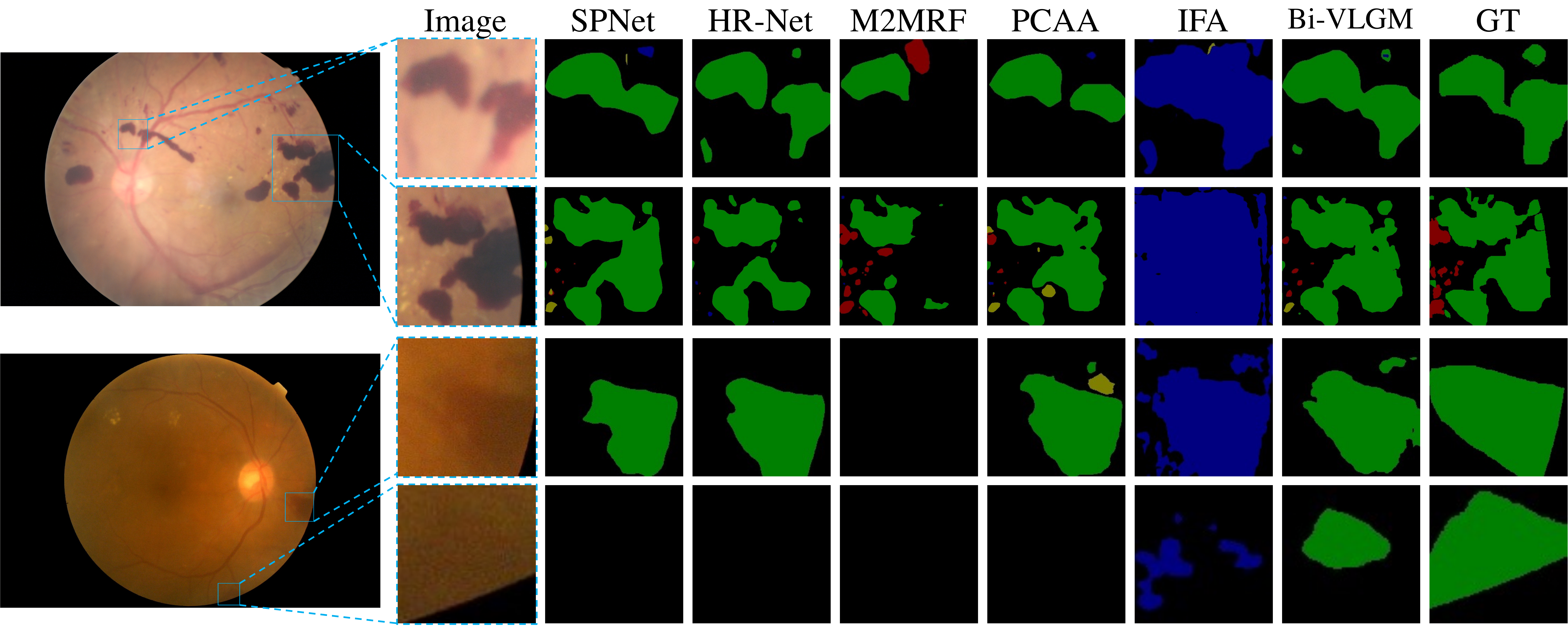}
   \caption{The visual results on the DDR dataset in comparison with the state-of-the-art methods.}
   \label{fig:vis_ddr}
\end{figure}

As visualized in Fig. \ref{fig:vis_ddr}, we compare our method against existing methods. It is observed that the HE regions of SPNet, M2MRF, and PCAA are much coarser compared with our Bi-VLGM in the first row. 
Moreover, in the last two rows, some inconspicuous HE regions are ignored by current methods, while our Bi-VLGM can recognize them. This reveals that the proposed Bi-VLGM can capture both the obvious and ambiguous features for different lesion classes to achieve better visual results than other methods.

\subsection{Ablation Study}

\paragraph{Ablation Experiments.} 
 We conduct detailed ablation experiments on IDRiD dataset to evaluate the effectiveness of word-level VLGM (Word) and sentence-level VLGM (Sentence) modules. The HR-NetV2 \cite{wang2020deep} is adopted as our baseline model. 
As illustrated in Table \ref{ablation}, word-level VLGM module significantly improves the performance of the baseline model by 2.26\% mIoU, indicating the effectiveness of word-level VLGM module. When further integrating sentence-level VLGM module, the segmentation performance is increased by 1.37\% mIoU, implying that sentence-level VLGM module is effective for segmentation.

\paragraph{Effectiveness of Severity-Aware Prompting.} We evaluate the effectiveness of severity-aware prompting by using different numbers of medical prompts in our Bi-VLGM. As listed in Table \ref{prompt}, the severity-aware prompting with 1 template achieves the worst segmentation performance with mIoU of 50.51\%.  When we adopt more than one template, the segmentation performance is improved remarkably by more than 0.86\% mIoU and arrives at 51.37\% mIoU with five templates. It indicates that larger diversity of medical prompts is more effective to extract severity features for segmentation.

\paragraph{Effectiveness of VLGM.} The proposed Bi-VLGM introduces VLGM to reformulate the vision-language matching as graph matching problem. To evaluate the effectiveness of the VLGM, we compare the performance when replacing the graph matching with the contrastive loss. To replace graph matching with contrastive loss, in the word level, we consider the local-class features pairs from the same lesion class as positive pairs and others as negative pairs. In the sentence level, we regard the global-severity features pairs as positive pairs if originating from the same image, and otherwise as negative pairs. As demonstrated in Table \ref{contrastive}, VLGM outperforms contrastive loss by a large margin of about 2\% mIoU, which suggests the effectiveness of our proposed VLGM. 

\begin{figure}[t]
  \centering
   \includegraphics[width=\linewidth]{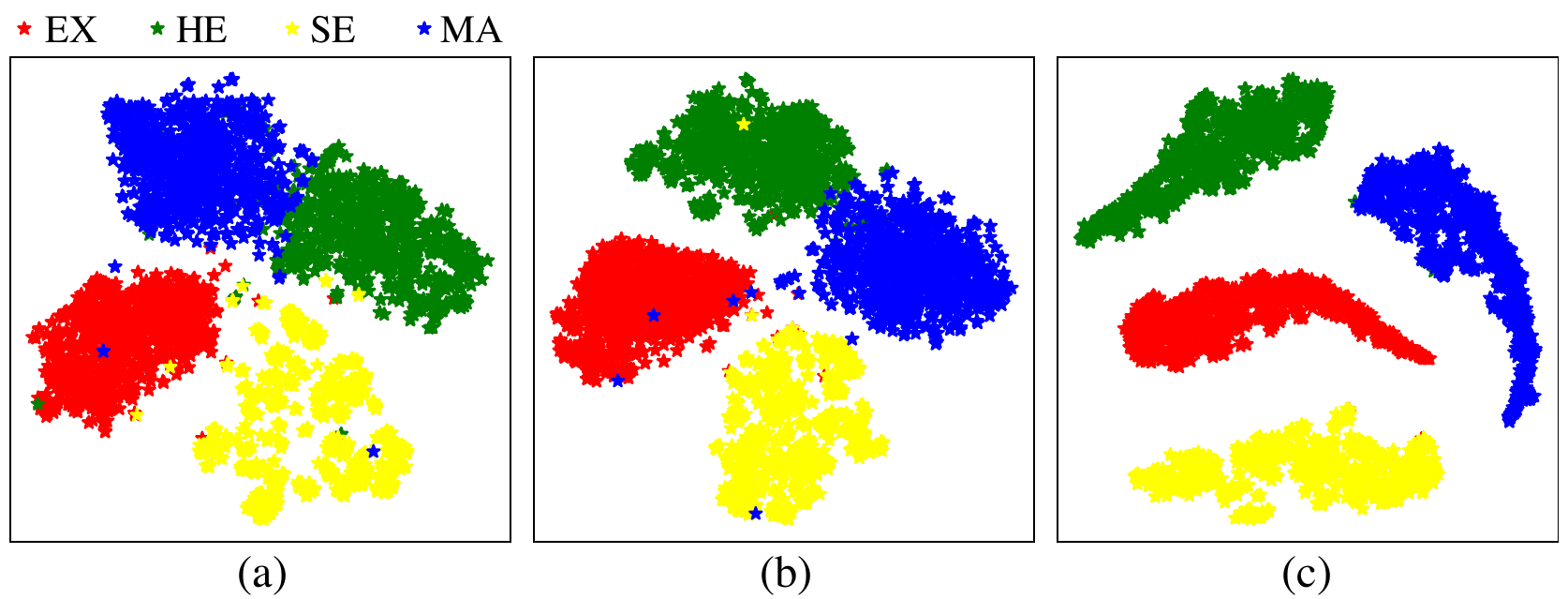}
   \caption{Feature comparison via T-SNE among (a) the baseline model, (b) the Bi-VLGM with contrastive loss and (c) the proposed Bi-VLGM.}
   \label{fig:vis_tsne}
\end{figure}

To compare the consistency of the high-order relation, we perform feature comparison via T-SNE among the baseline model, Bi-VLGM and its contrastive loss version. For each lesion class, we randomly sample 1000 pixels on the last hidden features of the segmentation model and present the T-SNE comparison in Fig. \ref{fig:vis_tsne}. The relation among each class in the baseline model should be remained after VLM. For instance, EX and SE, HE and MA, and EX and MA are close to each other due to their similar features \cite{jaya2015detection, das2022critical,alghadyan2011diabetic}. However, Bi-VLGM with contrastive loss distorts the relation between EX and MA, making their distance larger. In contrast, Bi-VLGM can preserve all the relations of the baseline model and also make the features more compact, implying the effectiveness of VLGM to preserve the intra-modal relation.

\begin{table}[t]
	\centering
	\caption{Comparison of VLGM and contrastive loss.}

\scalebox{0.9}{
		\begin{tabular}{p{2.5cm}<{\centering}|p{1.2cm}<{\centering}|p{1.2cm}<{\centering}|p{1.2cm}<{\centering}}
			\toprule[1pt]
 Methods & IoU & F1 & AUPR \\ \hline
Contrastive Loss & 49.46 & 65.08 & 66.23 \\ 
VLGM & \textbf{51.37} &  \textbf{66.71} &  \textbf{68.12} \\ 
			\bottomrule[1pt]
	\end{tabular}}
	\label{contrastive}
\end{table}

\section{Conclusion}
\label{sec:conclude}
In this paper, we introduce a Bi-level class-severity-aware Vision-Language Graph Matching (Bi-VLGM) for text guided medical image segmentation, consisting of a word-level VLGM module and a sentence-level VLGM module. It aims to exploit the relation of the local region and lesion class, and the relation of global image and disease severity level. In word-level VLGM, to remain the intra-modal relation consistent, we introduce a vision-language graph matching (VLGM) to reformulate VLM as graph matching problem and perform VLGM between local region and class-aware prompts to bridge their gap. In sentence-level VLGM, to provide disease severity information, we introduce a severity-aware prompting to quantify the severity level of retinal lesion, and mine the relation between the global image and the severity-aware prompts. By investigating the relation between the local (global) and class (severity) features, the segmentation model can selectively learn the class-aware and severity-aware information to promote the performance.
Extensive experiments prove the effectiveness of our method and our superiority to existing methods.


{\small
\bibliographystyle{ieee_fullname}
\bibliography{egbib}
}

\end{document}